\documentstyle[epsfig,times]{aa}

\def\B{{\em BeppoSAX\/}}

\def\ergcms{\mbox{erg cm$^{-2}$ s$^{-1}$}}
\def\@cite#1#2{(#1\if@tempswa , #2\fi)}
\title{Hard X--rays from Type II bursts of the Rapid Burster and its 
transition toward quiescence}

\author{N. Masetti\inst{1}
\and
F. Frontera\inst{1,2}
\and
L. Stella\inst{3,4}
\and
M. Orlandini\inst{1}
\and
A.N. Parmar\inst{5}
\and
S. Del Sordo\inst{6}
\and
L. Amati\inst{1}
\and
E. Palazzi\inst{1}
\and
D. Dal Fiume\inst{1}
\and
G. Cusumano\inst{6}
\and
G. Pareschi\inst{7}
\and
I. Lapidus\inst{8}
\and
R.A. Remillard\inst{9}
}

\institute{Istituto Tecnologie e Studio delle Radiazioni Extraterrestri,
CNR, via Gobetti 101, I-40129 Bologna, Italy
\and
Dipartimento di Fisica, Universit\`a di Ferrara, via Paradiso 11, I-44100
Ferrara, Italy
\and
Osservatorio Astronomico di Roma, via Frascati 33, I-00040 Monteporzio
Catone, Italy
\and
Affiliated to International Center for Relativistic Astrophysics
\and
Astrophysics Division, Space Science Department of ESA, ESTEC, Postbus
299, NL-2200 AG Noordwijk, The Netherlands
\and
Istituto di Fisica Cosmica ed Applicazioni all'Informatica, CNR, via
La Malfa 153, I-90146 Palermo, Italy
\and
Osservatorio Astronomico di Brera, via Bianchi 46, I-23807 Merate, Italy
\and
McKinsey \& Co., Inc., No. 1 Jermyn Street, London SW1Y 4UH, United
Kingdom
\and
Center for Space Research and Department of Physics, Massachusetts
Institute of Technology, 77 Massachusetts Avenue, Cambridge, MA 02139,
USA
}

\offprints{N. Masetti, masetti@tesre.bo.cnr.it}
\date{Received 13 June 2000; Accepted 29 August 2000}
\thesaurus{06          % Form. Struct. and Evolution of Stars
           (08.09.2: MXB\thinspace1730-335;
            08.14.1;   % Stars: neutron
            13.25.1;   % X--rays: bursts
            13.25.3;   % X--rays: general
            13.25.5)   % X--rays: stars
}

\begin{document}

% This will write DRAFT across each page
%\special{!userdict begin /bop-hook{gsave 200 30 translate
%65 rotate /Times-Roman findfont 220 scalefont setfont
%0 0 moveto 0.80 setgray (DRAFT) show grestore}def end}

\maketitle

\markboth{N. Masetti et al.: Hard X-rays from the Rapid
Burster and its decay to quiescence}{N. Masetti et al.: 
Hard X-rays from the Rapid Burster and its decay to quiescence}

\begin{abstract}

We report on 4 \B\ Target Of Opportunity observations of
MXB\thinspace1730--335, the Rapid Burster (RB), made
during the 1998 February--March outburst. In the first observation,
approximately 20 days after the outburst peak, the X--ray light curve
showed Type II bursts at a rate of 43 hr$^{-1}$.
Nine days later, during the second \B\ pointing, only 5 Type II bursts
were detected at the beginning of the observation.
During the third pointing no X--ray bursts were detected and in the
fourth and final observation the RB was not detected at all. Persistent
emission from the RB was detected up to 10 keV during the first three
pointings.
The spectra of the persistent and bursting emissions below 10 keV were
best fit with a model consisting of two blackbodies. An additional
component (a power law) was needed to describe the 1--100~keV
bursting spectrum when the persistent emission was subtracted.
To our knowledge, this is the first detection of the RB 
beyond 20~keV. We discuss the evolution of the spectral parameters
for the bursting and persistent emission during the outburst decay.
The light curve, after the second \B\ pointing, showed a steepening of
the previous decay trend, and a sharper decay rate leading to
quiescence was observed with \B\ in the two subsequent observations. We
interpret this behaviour as caused by the onset of the propeller effect.
Finally, we infer a neutron star magnetic field $B\sim4 \times 10^8$~Gauss.


\end{abstract}

\section{Introduction}
\label{intro}
The Rapid Burster (MXB\thinspace1730--335; hereafter RB) is a well
known and extensively studied low mass X--ray binary (LMXRB; see the
recent review by
Lewin et al. 1995)\nocite{Lewin95} located in the globular cluster Liller
1 at a distance of $\sim$8~kpc (\cite{Ortolani96}). 
It is a recurrent transient with outbursts which last for few weeks
followed by quiescent intervals which generally last for $\sim$6 months.
Recently, a likely radio counterpart to the RB was observed by Moore et
al. (2000)\nocite{Moore00} during several X--ray outbursts.

The RB is unique among LMXRBs in that it displays two 
different kinds of X--ray bursts: Type I bursts, which are typical of 
many LMXRBs with low-magnetic field neutron stars (NS) and Type II bursts.
The former events are interpreted as due to nuclear burning of accreted
material onto the surface of the NS; Type II bursts, instead, probably
result from spasmodic accretion onto the NS surface (e.g. Lewin et al.
1995). Until 1996, the RB was the
only X--ray source from which Type II events were detected. Now another
transient X--ray source, GRO J1744--28, is known to show hard X--ray
bursts, that likely have the same origin of the
RB Type II bursts (\cite{Kouveliotou96}). The NS nature of GRO J1744--28
has been clearly established from the observation in its X--ray
emission of coherent 0.467~s pulsations (\cite{Finger96}). 

The RB shows a variety of emission modes. At some times it appears as a
typical low magnetic field LMXRB with persistent emission (PE) and Type I
bursting emission (BE);
but there are times when both Type I and Type II BE are observed, and
times with only Type II BE. In the presence of
Type II bursts only, the PE is well visible after long 
($\ga$30~s) events, but is weak or absent during short Type II events. 
The time behaviour of the PE between Type II bursts, when
it is observable, is complex with dips (before and after Type II burts),
humps, glitches, bumps (\cite{Lubin93}) and periodicities known as
`ringing tail' observed shortly after long Type II bursts
(\cite{Lubin92}).

Type II bursts have been observed with durations between $\sim$3~s and
640~s: the shortest events
(less than $\sim$20~s) show a sharp rise and a decay with multiple peaks
(`ringing'), instead the longer events are flat topped. Their time
behaviour is like that of a relaxation oscillator: their fluence $E$ is
roughly proportional to the time interval, $\Delta t$, to the following
burst (\cite{Lewin76}).

Quasi-periodic oscillations (QPOs) with centroid frequencies in the range
from $\sim$2 to $\sim$7~Hz were observed in Type II bursts; the 
centroid frequency was found to be anti-correlated with the burst peak
flux (see review by Lewin et al. 1995)\nocite{Lewin95}. Also in the PE
QPOs were observed with centroid frequencies from $\sim$0.04 
to $\sim$4.5~Hz with no frequency--intensity correlation, a repetitive
pattern from $\sim$4~Hz to to $\sim$2~Hz, and a 
positive correlation with spectral hardness (\cite{Stella88}).

The spectral properties of the Type II BE and of the PE have been
extensively investigated in the X--ray
energy band up to about 20~keV. Some authors (\cite{Marshall79};
\cite{Barr87}) found
that Type II burst spectra were well fit by a blackbody (BB) spectrum
with $kT\sim$ 1.5--2~keV, while Stella et al. (1988)\nocite{Stella88},
analyzing long Type II bursts, found that the best fit was
obtained with an unsaturated Comptonization model ($N(E)\propto 
E^{-\Gamma} exp(-E/kT)$, with $\Gamma\sim0.7$ and $kT\sim2.6$~keV for
1.5-2 min events). Instead, a two-component blackbody (2BB) plus a power
law (PL) provided the best fit to the spectra between 2 and
20~keV (\cite{Guerriero99}).
The unsaturated Comptonization model was found to describe well the PE
spectra (\cite{Barr87}; \cite{Stella88}), with $\Gamma\sim-0.06$ and
$kT\sim2.8$~keV 
in the time periods after 1.5-2 min Type II events (\cite{Stella88}).
Instead Guerriero et al. (1999)\nocite{Guerriero99} found that the PE is well
described by the same model used for the Type II bursts (i.e. 2BB+PL). 
An observation of the source in quiescence
with the ASCA satellite provided a positive detection with a 2--10 keV
luminosity of (1.9$^{+1.3}_{-0.6})\times 10^{33}$~erg s$^{-1}$, for a
distance of 8~kpc and an hydrogen column density of 1$\times
10^{22}$~cm$^{-2}$ (Asai et al. 1996).

Up to now, a detailed study of the evolution of the spectral
properties and the decay profile to quiescence of the RB has never been
performed. In addition, the high-energy emission from the source is still
poorly known (see, e.g., \cite{Claret94}). In order to address these
issues, a series of observations was performed during the 1998
February-March outburst (Fox et al. 1998) with the \B\ satellite
(\cite{Boella97a}).
In Sect.~2 we describe the observations, in Sect.~3 we present
the spectral results, while in Sect.~4 we discuss them and the source
transition to quiescence together with their implications.

\begin{figure}
\vspace{-1cm}
\epsfig{figure=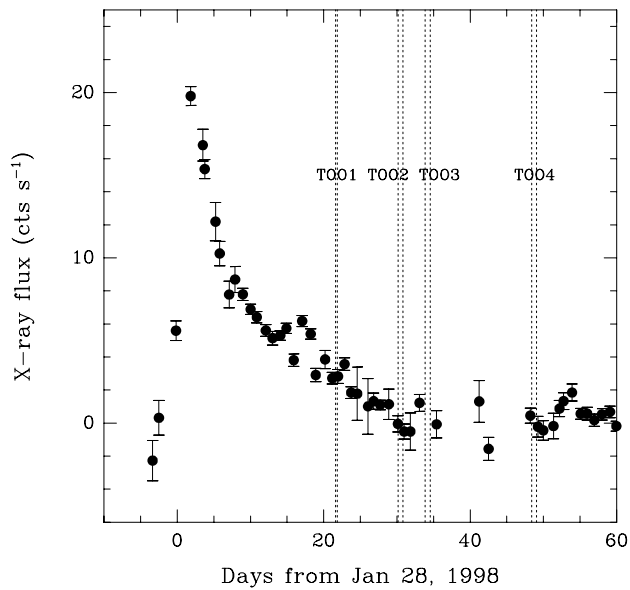,width=10cm}
\vspace{-1.5cm}
\caption[]{{\it R-XTE}/ASM 2--10~keV light curve of the 1998
February-March outburst of the RB. The vertical dashed lines indicate the
times of the 4 {\it BeppoSAX} TOO observations}
\end{figure}

\begin{figure*}
\vspace{-5cm}
\begin{center}
\epsfig{figure=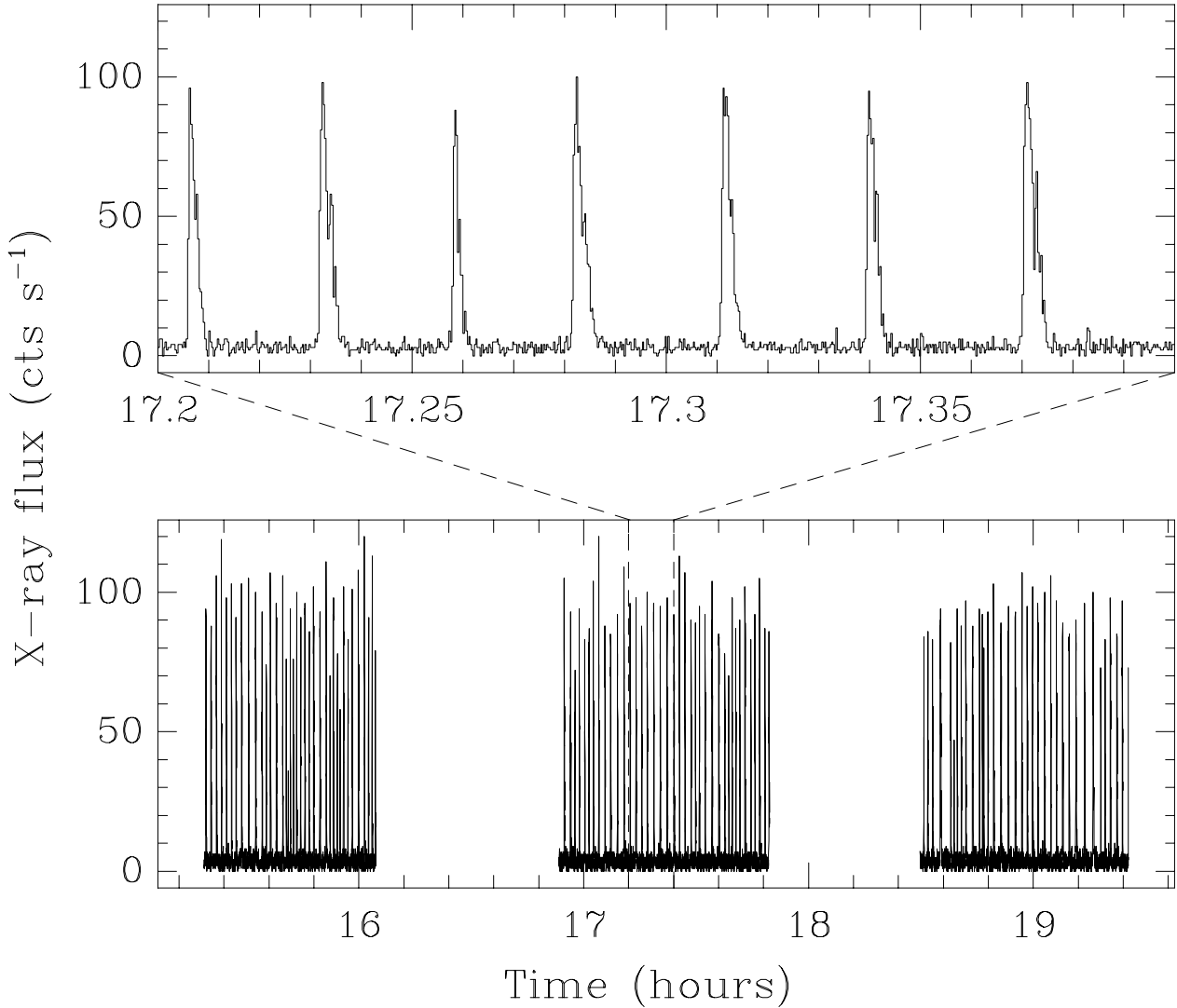,width=19cm}
\end{center}
\vspace{-6cm}
\caption[]{MECS 2--10~keV light curve of TOO1. Times are
expressed in hours from 0 UT of 1998 February 18. On top of the Figure,
the enlargement of the central part of the light curve is shown}
\end{figure*}

\begin{figure*}
\vspace{-5cm}
\begin{center}
\epsfig{figure=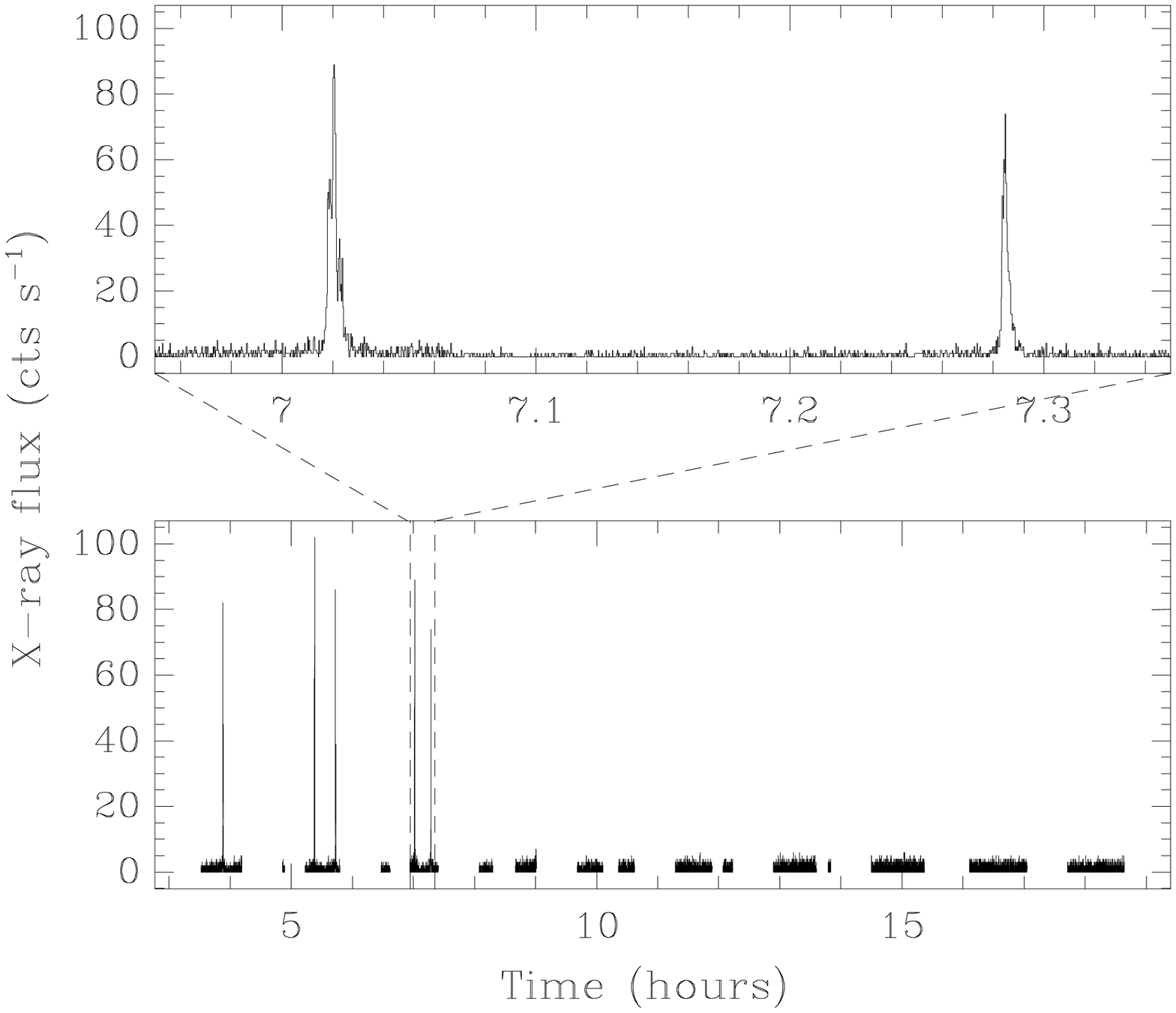,width=19cm}
\end{center}
\vspace{-6cm}
\caption[]{MECS 2--10~keV light curve of TOO2. Times are expressed
in hours from 0 UT of 1998 February 27. Note the marked reduction in
bursting activity with respect to TOO1 (Fig. 2). On top of the Figure,
the enlargement of the section of the TOO2 light curve containing the last
two bursts is shown}
\end{figure*}

\section{Observations}
\label{obs}
The RB was observed with the \B\ Narrow Field Instruments (NFIs) four times
as a Target Of Opportunity (TOO).
The NFIs include the Low-Energy Concentrator
Spectrometer (LECS: 0.1--10~keV; \cite{Parmar97}), two Medium-Energy
Concentrator Spectrometers (MECS: 1.5--10~keV; \cite{Boella97b}), 
and the Phoswich Detection System (PDS: 15--300~keV;
\cite{Frontera97}). The two MECS consist of grazing incidence telescopes
with imaging gas scintillation proportional
counters in their focal planes. The LECS uses an identical concentrator
system as the MECS, but utilizes an ultra-thin entrance window and
a driftless configuration to extend the low-energy response down to
0.1~keV. The LECS and the MECS have a circular field of view with
diameter 37$'$ and 56$'$, respectively.
The PDS (field of view of
$\sim$1$^{\circ}$.3$\times$$\sim$1$^{\circ}$.3 FWHM) consists
of four independent detection units arranged
in pairs, each one having a separate rocking collimator. 

Table~1 reports the log of the TOO observations presented in this paper.
They spanned just over one month (from February 18 to March
18) and caught the object during different emission modes and flux levels
as it was decaying from an X--ray intensity of about 20\% that of the
outburst peak down to quiescence.
Figure 1 shows the 2--10~keV light curve obtained with the {\it All-Sky
Monitor} (ASM) onboard the {\it Rossi X--ray Timing Explorer} ({\it R-XTE})
satellite\footnote{{\it R-XTE}/ASM light curves of X--ray
sources are available at {\tt http://space.mit.edu/XTE/asmlc/}~.}.
Also shown in Fig. 1 are the times of the four \B\ observations.

Due to the
presence of a strong and variable X--ray source, 4U\thinspace1728--34 
(=GX\thinspace354--0, located $\sim$30$\arcmin$ away from the RB), a
special observation strategy for the latter instrument was adopted in
order to minimize the contamination from this source: the PDS rocking
collimators were offset by 40$\arcmin$ along a direction opposite from
that of 4U\thinspace1728--34.
Unfortunately, due to an incorrect instrumental setting, these
collimators did not move during TOO2 and TOO3. Thus only the LECS  
and MECS data are usable for these two observations.

During the last observation the RB, very faint at that time, was not detected 
with the NFIs. Contamination from 4U~1728--34, which was unusually intense at 
the time, adversely affected the sensitivity of all the NFIs during this
observation. This issue is analyzed in more detail in Sect. 3.4 of this
paper.

\begin{figure}
\vspace{-5cm}
\begin{center}
\epsfig{figure=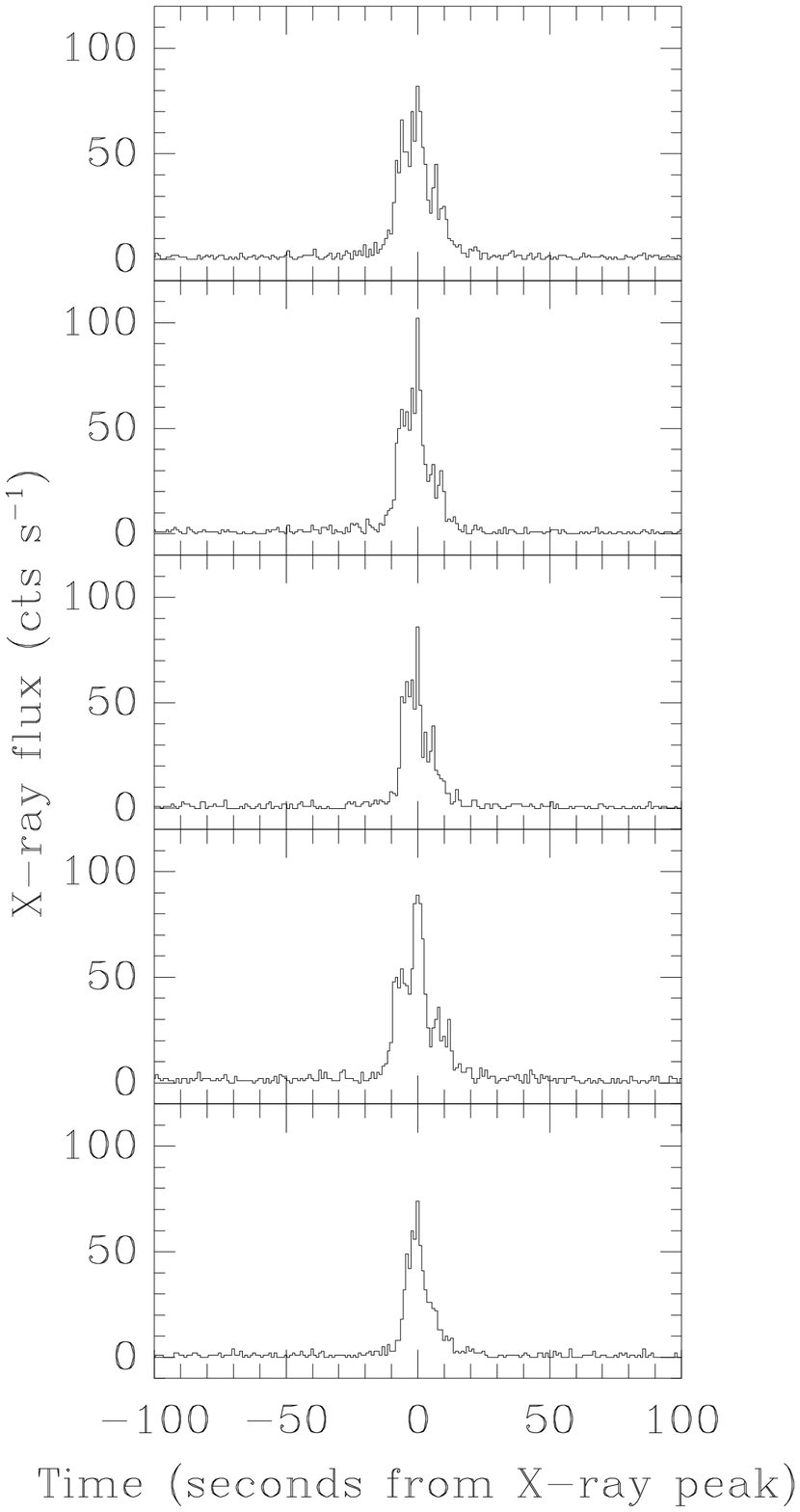,width=15cm}
\end{center}
\vspace{0.3cm}
\caption[]{2--10~keV band light curves of the 5 bursts observed during
TOO2. Times are expressed in seconds from the burst peak. Different shapes
and lengths with respect to TOO1 bursts (see enlargement of Fig. 2) are
apparent}
\end{figure}

\begin{table*}
\caption[]{{\it BeppoSAX} observation log of the TOOs presented in this
paper}
\begin{center}
\begin{tabular}{ccccccc}
\noalign{\smallskip}
\hline
TOO & Outburst day & Start day & Start time & Duration & Exposure &
2--10 keV MECS\\
    &              &  (1998)   & (UT)       & (ks)     & (MECS; ks) &
count rate (s$^{-1}$)\\
\noalign{\smallskip}
\hline
\noalign{\smallskip}
1 & 21 & \multicolumn{1}{l}{Feb 18} & 15:17& 17.5 & 9.5 & 7.88\\
2 & 30 & \multicolumn{1}{l}{Feb 27} & 03:55 & 57.1 & 27.0 & 1.01\\
3 & 33 & \multicolumn{1}{l}{Mar 2} & 19:57 & 61.5 & 29.8 & 0.15\\
4 & 48 & \multicolumn{1}{l}{Mar 17} & 09:52 & 56.7 & 31.3 &\llap{$<$}0.02\\
\noalign{\smallskip}
\hline
\noalign{\smallskip}
\end{tabular}
\end{center}
\end{table*}

\section{Data analysis and results}
\label{results}

Good NFI data were selected from intervals when the elevation angle
above the Earth limb was $>$$5^{\circ}$ and when the
instrument functioning was nominal. The SAXDAS 2.0.0 data
analysis package (\cite{Lammers97}) was used for the LECS and MECS.
The PDS data reduction was performed using XAS version 2.1
(\cite{Chiappetti97}).
The LECS and MECS events from the RB were extracted
from circular regions with radii between 3\arcmin\ and 8\arcmin, centred
on the source position. These exctraction radii were chosen case by case in 
order to optimize the signal-to-noise ratio for different X-ray intensities 
of the RB. The MECS exposure times are given in Table~1.
Background subtraction for the two imaging instruments
was performed using standard files, while the background for the PDS data
was evaluated from offset fields.

The 2--10 keV MECS light curves of TOO1 and TOO2, with a time binning of
1~s, are shown in Fig.~2 and Fig.~3, respectively. The gaps in the light
curves are due to non-observing intervals during Earth occultations and
during passages through the South Atlantic Geomagnetic Anomaly.
The mean source count rates with the MECS are given in Table~1.

The LECS and MECS spectra were rebinned to oversample by a factor 3
of the FWHM of the energy resolution, and having a
minimum of 20 counts per bin such that the $\chi^2$ statistics could
reliably be used.
The PDS spectra were rebinned using the standard techniques in SAXDAS.
Data were selected in the energy ranges where the instrument responses
were well determined: 
0.5--8.0~keV for the LECS, 1.8--10~keV for the MECS, and 15--100~keV for
the PDS.
The only exception to this choice was during TOO1, in which
LECS data were selected between 1 and 8~keV due to poor signal-to-noise
ratio below 1~keV. We used the package {\sc xspec} v10.0
(\cite{Arnaud96}) for the spectral fitting.
In the broadband fits, normalization factors were applied to the LECS and
PDS spectra following the cross-calibration tests between these
instruments and the MECS (\cite{Fiore99}).
For clarity of display, the spectra from multi-instrument fits (Figs.
5, 6 and 7) are shown normalized to the level of the MECS.
The reported uncertainties throughout the paper are single parameter
errors at 90\% confidence level.
For the luminosity estimates we will assume a distance to the RB of
8 kpc (Ortolani et al. 1996).

\begin{table*}
\caption[]{Best-fit parameters for the TOO spectra.
$L$ is the unabsorbed luminosity in units of 
$10^{36}$~erg~s$^{-1}$ assuming a distance of 8~kpc}
\begin{center}
\begin{tabular}{c|ccc|cc|c}
\hline
\noalign{\smallskip}
 & \multicolumn{3}{c|}{TOO1} & \multicolumn{2}{c|}{TOO2} & TOO3 \\
\noalign{\smallskip}
\hline
\noalign{\smallskip}
Model and & 1--10 keV & 1--10 keV & 1--100~keV &
1.8--10 keV & 0.5--10 keV & 0.5--10 keV \\
parameter & BE & PE & BE $-$ PE & BE & PE & PE\\
\noalign{\smallskip}
\hline
\noalign{\smallskip}
${\chi^{2}}$/$\nu$ & 463/395 & 217/203 & 262/223 & 59/128 & 110/113 &
59/55\\
\multicolumn{1}{l|}{2BB:} & & & & &\\
$N_{\rm H}$ ($\times$10$^{22}$ cm$^{-2}$) & $3.5\pm0.5$ & $1.5\pm0.3$ &
$4.9^{+1.7}_{-1.1}$& [4.0$^{\rm a}$] & $1.6\pm0.3$ & $1.1^{+0.4}_{-0.3}$\\
$kT_1$ (keV) & $0.46\pm0.04$ & $0.63\pm0.06$ &
$0.32^{+0.23}_{-0.12}$ & 0.46$^{+0.10}_{-0.08}$ & $0.65\pm0.07$ &
$0.64\pm0.09$ \\
$L^{\rm BB}_1$ & $15^{+6}_{-5}$& $0.85^{+0.08}_{-0.06}$ &
$10^{+22}_{-10}$ & $14^{+4}_{-3}$ & $0.32^{+0.04}_{-0.03}$ &
$0.06\pm0.01$ \\
$R_1^{\rm BB}$~(km) & $51\pm13$ & 6.5$\pm$1.3 &
$90^{+160}_{-80}$ & $48^{+32}_{-19}$ & 3.7$\pm$0.8 & 1.7$\pm$0.5 \\
$kT_2$ (keV)& $1.64\pm0.03$ & $1.72^{+0.19}_{-0.17}$ &
$1.67\pm0.04$ & $1.57^{+0.19}_{-0.13}$ & $1.78^{+0.10}_{-0.08}$ &
$2.1^{+0.4}_{-0.2}$ \\
$L^{\rm BB}_2$ & $47\pm0.8$ & $0.74^{+0.06}_{-0.05}$ &
$40\pm2$ & $24.2^{+1.4}_{-1.6}$ & $0.76\pm0.03$ &
$0.106^{+0.011}_{-0.008}$ \\
$R_2^{\rm BB}$~(km) & $7.1\pm0.7$ & $0.81\pm0.17$ &
$6.3 \pm 0.3$ & 5.6$^{+1.2}_{-1.3}$ & 0.76$\pm$0.08 &
0.21$^{+0.07}_{-0.04}$ \\
\multicolumn{1}{l|}{+ power law:} & & & & & & \\
$\Gamma$ & & & $3.1^{+0.3}_{-0.4}$& & & \\
$K$ (ph keV$^{-1}$ cm$^{-2}$ s$^{-1}$ at 1 keV) & & & $3.3^{+4.5}_{-2.5}$
& & & \\
\multicolumn{1}{l|}{+ Fe emission line:} & & & & & & \\
$E_{\rm l}$ (keV) & & & $6.5 \pm 0.2$ & & & \\
EW (eV) & & & $100^{+80}_{-60}$& & & \\
FWHM (keV) & & & $0.4^{+0.3}_{-0.4}$& & & \\
$I_{\rm l}$ ($\times 10^{-3}$ ph cm$^{-2}$ s$^{-1}$) & & &
$7^{+5}_{-4}$ & & & \\
\noalign{\smallskip}
\hline
\noalign{\smallskip}
L$_{\rm 2-10~keV}$ & $43.5\pm0.3$ & $1.08\pm0.02$ &
$47.3 \pm 0.3$ & $24.6 \pm 0.9$ & $0.78\pm0.01$ & $0.109\pm0.002$ \\
L$_{\rm 10-100~keV}$ & & & $7.2 \pm 1.1$ & & & \\
\noalign{\smallskip}
\hline
\noalign{\smallskip}
\multicolumn{7}{l}{$^{\rm a}$ fixed at the best fit value determined from
the 1--10~keV BE spectrum of TOO1} \\
\noalign{\smallskip}
\hline
\end{tabular}
\end{center}
\end{table*}

\subsection{TOO1}
\label{TOO1}
During TOO1, the RB was in a state of strong bursting activity: indeed,
the MECS light curve (Fig. 2) showed 113 X--ray bursts during 9.5~ks of
good observation time, corresponding to an average rate of 43 bursts
hr$^{-1}$. Their Type II character is apparent from the `ringing' visible
in their profiles and by the time between them. 
The RB appeared to be in the Phase II, Mode II of the classification by
Marshall et al. (1979)\nocite{Marshall79}: all the events occurred almost
regularly (at a time distance of about 100~s) and had short time
durations (about 15~s). Also in the 0.1--2~keV LECS data these Type II
bursts were visible. The total time-averaged 2--10~keV unabsorbed flux
level during TOO1 was 8.2$\times 10^{-10}$~\ergcms.

\subsubsection{1--10~keV spectrum of BE and PE}

For the spectral analysis in the 1--10~keV energy range, the MECS TOO1
data, binned at 1 s, were divided into two subsets: PE (below 5 counts
s$^{-1}$) and BE (above 5 counts s$^{-1}$). Then, LECS and MECS BE and PE
spectra were accumulated using these two time windows. Given the small
extraction radii used (4$\arcmin$ for the MECS, 8$\arcmin$ for the LECS),
the high RB intensity level and the low flux level of 4U\thinspace1728--34
at that time, the contamination from the latter source was negligible.

The BE 1--10~keV spectrum can be well fit ($\chi^{2}/\nu$ = 463/395,  
where $\nu$ indicates the number of degrees of freedom in the fit)
with a 2BB model, photoelectrically
absorbed (the hydrogen column density was modeled using the Wisconsin
cross sections as implemented in {\sc xspec}; see \cite{Morrison83}). The
same fitting model applies to the PE spectrum 
($\chi^{2}$/$\nu\,=\,217/203$). In Table~2 we show the fit results.
The 2BB model was also used by Guerriero et al. (1999)\nocite{Guerriero99}
to fit the {\it R-XTE}/PCA BE and PE spectra of the RB in the 2--20~keV
band. These authors however noted that the PE between 10 and 20 keV was
better fit by adding a PL component. In our 2--10~keV band spectra this
was not necessary; but, when the data beyond 10 keV were considered 
(see Sect. 3.1.2), an additional component had to be added.

As it can be seen from Table~2, the BB temperatures did not significantly
change for the PE and the BE, while the 2BB luminosities were much higher
during BE than during PE. The column densities appeared also to be
different for PE and BE spectra: $(1.5\pm 0.3)\times$10$^{22}$ cm$^{-2}$
and $(3.5\pm 0.5)\times$10$^{22}$ cm$^{-2}$, respectively. The increase of
the $N_{\rm H}$ column density during the BE is apparent.
The value obtained for the PE is consistent with the color excess $E(B-V)$
measured along the RB direction (\cite{Tan91}).

From the MECS 2--10~keV TOO1 observation we also computed the ratio,
$\alpha$, between the (unabsorbed) PE and BE fluences integrated over the
TOO1 observation time. We found that $\alpha$ = 0.154 $\pm$0.002, that is
about two orders of magnitude lower than the mimimum value observed 
($\sim$10) in the case of Type I X--ray bursts 
(\cite{Lewin95}). This result confirms the Type II character of the
detected bursts.

\subsubsection{1--100~keV spectrum of the BE}

During TOO1 the RB was also visible in hard X--rays (15--100 keV).
However, due to the lower statistical quality of the data, 
the PDS light curve was much noisier and it was
difficult to single out the bursts. In order to separate the
BE from the PE, two PDS spectra (`persistent' and `bursting') were
accumulated using the time intervals of the bursts given by the MECS data.
Given the residual contamination from 4U\thinspace1728--34, we only derived
the high energy spectrum of the BE using as background level the
count rate spectrum measured during the PE time intervals.

The combined LECS+MECS+PDS PE-subtracted BE spectrum in the
1--100~keV energy band is shown in Fig.~5.
As it can be seen, the source was clearly detected at high energies.
To our knowledge, this is the first time the RB is detected above 20~keV.
A simple 2BB model did not clearly fit the data
($\chi^{2}/\nu = 326/228$), especially at high energies ($>$10 keV).
A satisfactory fit ($\chi^{2}/\nu = 281/226$) was obtained by adding
to the 2BB model a PL. The best fit parameters are given in Table~2. 

We also tried to use, alternatively to the PL, a Comptonization model of
soft photons in a hot thermal plasma (Titarchuk 1994; {\sc CompTT} model
in {\sc xspec}), by assuming seed photons with temperature given by one of
the two temperatures of the 2BB model.
A fit quality ($\chi^{2}/\nu = 278/225$) similar to that achieved with
the PL model was obtained, but electron temperature and optical
depth of the comptonizing cloud were not constrained by the data
($kT_e\sim 20$~keV and $\tau \sim 0.4$, with large uncertianties).
When the Comptonization model was used, the fit quality did not depend
on the chosen seed photon temperature: both the cooler and the hotter BB
photons provided fits of comparable quality. Also, disk and spherical
geometries for the Comptonization resulted in equivalent fits.

The addition of a Fe K emission line to the 2BB+PL  
improved the fit ($\chi^2/\nu  =  262/223$); the probability of a
chance improvement, computed by means of an F-test, was quite low
($\sim$1$\times$10$^{-3}$). The flux and energy centroid derived
for the Fe K line (see Table~2) are consistent with the findings by Stella
et al. (1988)\nocite{Stella88}, while the width is lower by a factor
$\sim$3. However Barr et al. (1987) gave an upper
limit to the width of this line which is consistent with our measurement.
No evidence of the same line was found in the 1--10 keV PE spectrum; however
the flux was too low to allow a similar detection.

\subsection{TOO2}

During this observation the RB drastically
reduced its bursting activity (see Fig.~3): only 5 bursts were detected
at the beginning of the observation, with time duration of 
$\sim$30 -- 40 s
and time distance among them spanning from $\sim$20 to $\sim$90 minutes.
The general shape of these 5 bursts (Fig. 4) is similar each other and
differs from those of TOO1 for the pre-maximum during their rise. 
This appears to be a sort of `hybrid' shape
between short ($<$15 s) and long (1 min or more), flat-topped bursts (see,
e.g., Lubin et al. 1992, 1993, and Guerriero et al. 1999).

The 2--10~keV flux level of the PE decreased by a factor 1.4, while the
total time-averaged 2--10~keV unabsorbed flux ($1.0 \times
10^{-10}$~\ergcms) decreased by a factor of about 8 with respect to TOO1.

The 2--10~keV $\alpha$ ratio between the (unabsorbed) PE and BE fluences
integrated over the TOO2 observation time was 4.17 $\pm$ 0.15, thus
significantly higher than that found during TOO1,
but lower than the minimum value measured in the case of Type I X--ray
bursts. Given that some bursts might have been lost during the
source occultation by the Earth, we have also computed $\alpha$ for
the first three orbits of TOO2, i.e. where the 5 bursts were seen.
In this case, we found that $\alpha$ = 1.40 $\pm$ 0.05. This value is
still higher than that found in TOO1, but lower than the minimum value
measured in the case of Type I bursts. 

The 1.8--10~keV spectrum of the BE in TOO2 was still consistent with the
2BB
model with $kT_1 \sim 0.5$~keV and $kT_2 \sim 1.6$~keV, even though, due
to the poor statistics particularly in the LECS and below 2 keV, we could
not derive accurate parameter values. Therefore, to better constrain the
main parameters of the model, we fixed the $N_{\rm H}$ value to that
obtained from the spectral fit of the TOO1 1--10 keV BE.
The results of the TOO2 BE fit are reported in Table 2.

The PE also (see Fig. 6) was best fit ($\chi^2/\nu  =  110/113$) with
a photoelectrically absorbed 2BB model (see Table~2) with $kT_1\, =\,
0.65\pm0.07$~keV and $kT_2 \,= \, 1.78^{+0.10}_{-0.08}$~keV, although a
photoelectrically absorbed bremsstrahlung model also gave an acceptable
fit ($\chi^{2}/\nu = 122/115$). 
The 2BB temperatures of BE and PE did not appear to change significantly
from the corresponding values measured during TOO1. No substantial
improvement in the fit was seen if a PL or a Comptonization component was
included. No indication for a Fe K emission line was present in the TOO2
data.

\subsection{TOO3}

During TOO3, 33 days from the outburst onset, the source
further reduced its X--ray emission and no bursts were seen throughout the
observation. The unabsorbed and time-averaged 2--10~keV flux, $1.4\times
10^{-11}$~\ergcms, had decreased by a factor 7 with respect to TOO2.

The 0.5--10 keV spectrum of the PE is shown in Fig. 7. 
Again it was well fit ($\chi^2/\nu =  59/55$) with a photoelectrically
absorbed 2BB model (see Table ~2), with $kT_1 \, = \,0.64\pm0.09$~keV and 
$kT_2 \,= \,2.1^{+0.4}_{-0.2}$~keV. The temperature of the hotter component
seemed to be marginally higher than the corresponding value observed in
TOO2, while $kT_1$ was fully consistent with the value obtained during
TOO2. A photoelectrically absorbed bremsstrahlung gave a poorer fit
($\chi^2/\nu = 70/57$). As in the case of TOO2, no Fe K emission line was
detected.

\begin{figure*}
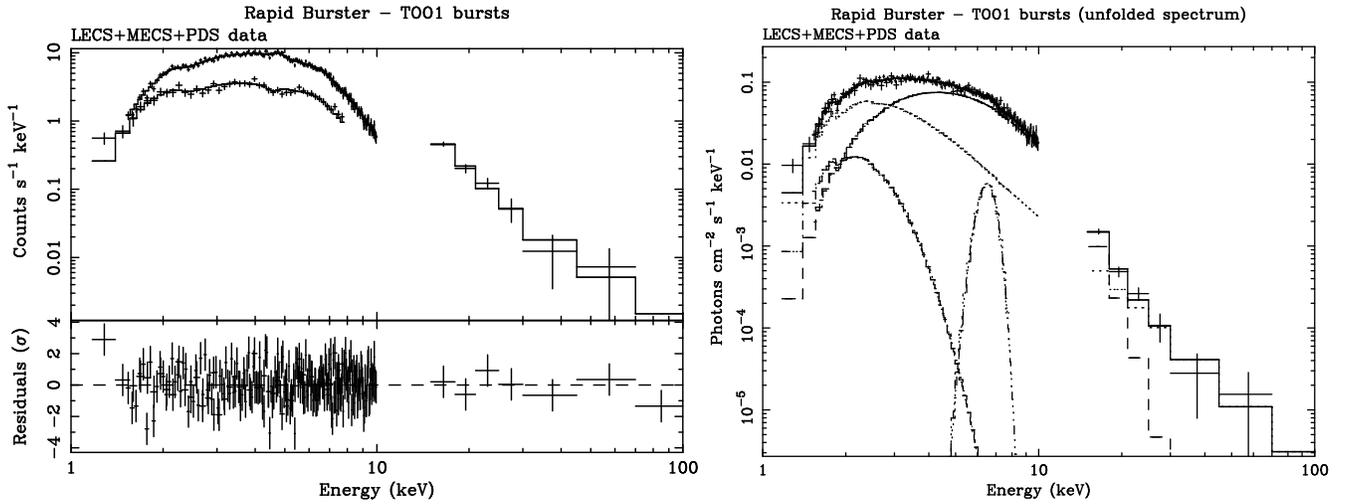

\begin{center}
\epsfig{figure=h2266f5l.ps,angle=-90,width=9.2cm}
\epsfig{figure=h2266f5r.ps,angle=-90,width=8.3cm}
\end{center}
\vspace{-.6cm}
\caption[]{TOO1 1--100 keV count rate (left panel) and photon
(right panel) spectra of the total PE-subtracted BE.
The fit corresponds to a photoelectrically absorbed two-component
blackbody plus power law model plus an iron line at 6.5 keV}
\end{figure*}

\begin{figure*}
\begin{center}
\epsfig{figure=h2266f6l.ps,angle=-90,width=9.2cm}
\epsfig{figure=h2266f6r.ps,angle=-90,width=8.3cm}
\end{center}
\vspace{-.6cm}
\caption[]{TOO2 0.5--10 keV count rate (left panel) and photon 
(right panel) spectra of the PE. The fit corresponds to a
photoelectrically absorbed two-component blackbody model}
\end{figure*}

\begin{figure*}
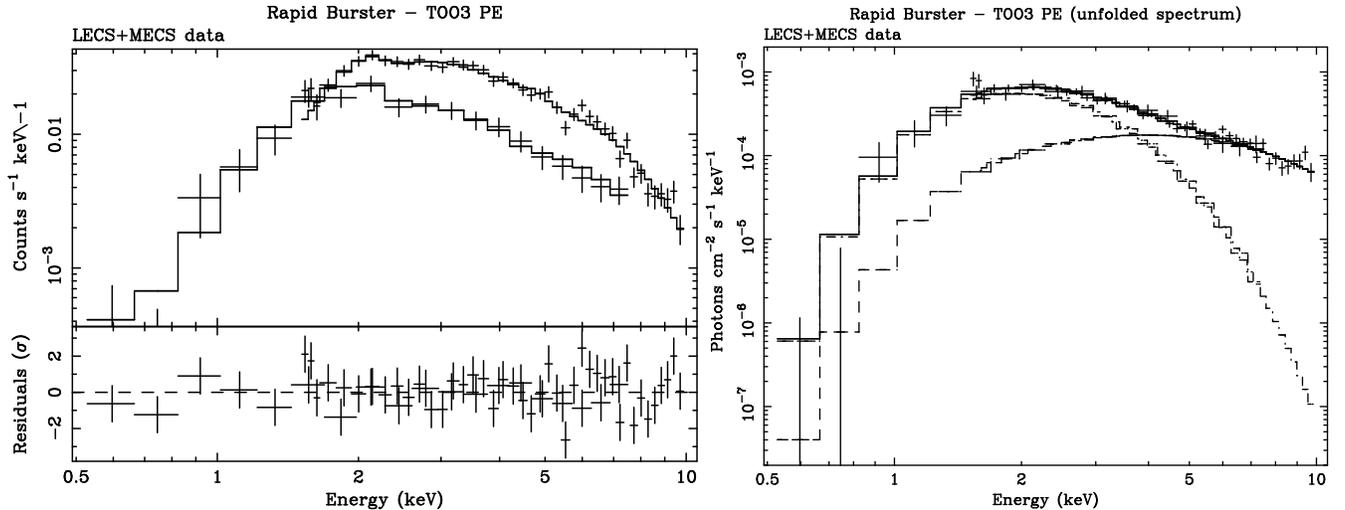

\begin{center}
\epsfig{figure=h2266f7l.ps,angle=-90,width=9.2cm}
\epsfig{figure=h2266f7r.ps,angle=-90,width=8.3cm}
\end{center}
\vspace{-.6cm}
\caption[]{TOO3 0.5--10 keV count rate (left panel) and photon 
(right panel) PE spectra. The fit corresponds to a
photoelectrically absorbed two-component blackbody model}
\end{figure*}

\subsection{TOO4}

During this observation the RB was no longer visible in the LECS and MECS
images. Instead, strong contamination due to stray light from
4U\thinspace1728--34 was apparent and extended up to the center of the
MECS image, where the RB emission was especially located.
4U\thinspace1728--34 was particularly active during TOO4 and prevented us
from getting a deep observation of the RB.
The PDS data were excluded from the analysis of TOO4 data since it proved
impossible to disentangle the RB emission from the contamination induced
by 4U\thinspace1728--34.

Using the background level present in the extraction circle of the RB
image, we evaluated the upper limit to the 2--10 keV count rate from the
RB. The result of this analysis indicates that, assuming the best fit
model spectrum of TOO3, an emission of 1.5$\times$10$^{-12}$ erg cm$^{-2}$
s$^{-1}$ from the RB would have been clearly detected at 3$\sigma$ at the
relevant position in the MECS field of view. Thus, we can conservatively
consider this value as a 3$\sigma$ upper limit to the RB X--ray emission
in the 2--10 keV energy band. The corresponding 3$\sigma$ upper limit to
the 2--10~keV luminosity is $1.1 \times 10^{34}$~erg s$^{-1}$. 

\section{Discussion}

The \B\ observations of the RB have permitted us to detect for the first time 
high energy X-rays ($>$20~keV) from Type II bursts and to study
the source in the late phase of the flux decay, from 21 to 48~days after 
the outburst onset. Type II bursts were detected in TOO1 and TOO2,
corresponding to 21 and 30 days since the outburst start, while the
PE was observed in all TOOs but the last observation, 
when the RB was not visible anymore.

\subsection{Spectral properties of the Type II bursts and of the PE}
\label{spect}

The broad band (1--100~keV) PE-subtracted spectrum of the BE,
measured during TOO1, is well described by a 2BB+PL model, 
with BB temperatures of $\sim$0.3 and $\sim$1.7~keV, while the photon
index of the high energy PL component is $\sim$3.  The presence of
a high energy component is not apparent in the 1--10~keV spectrum,
well described by a simple 2BB model. A thermal Comptonization model
(\cite{Titarchuk94}) also fits the high energy component of the BE
PE-subtracted spectrum, but the thermal plasma parameters are not well
constrained.

The presence of a high energy component is common in persistent
low-luminosity LMXRBs (mainly X-ray bursters, see e.g. the review paper by 
Tavani \& Barret 1997), and it is also seen during the outburst in some
Soft X-ray Transients (SXTs) containing NSs (e.g., Aql X-1 and Cen X-4,
see \cite{Campana98}). Instead this component is generally
not observed in high-luminosity LMXRBs (Z-sources; see e.g.
\cite{White88}). 
Only recently, thanks to \B, evidence of high energy components in Z 
sources has been reported from some objects
(\cite{Frontera98}; \cite{Masetti00}). This component has been interpreted
in these sources as due to the presence, along with an accretion disk, of a  
cloud of hot electrons which comptonize soft photons coming from the
disk, similarly to what occurs in stellar mass black-hole candidates 
(\cite{Barret00}).

The BE spectrum (not PE-subtracted) below 10~keV during TOO1 and TOO2 
is well fit with a 2BB model, in spite of a different shape
of the burst time profiles as seen in these two \B\ pointings.
The 2BB temperatures of the BE do not appear to change
from TOO1 to TOO2, as well as the cooler BB luminosity, while the
hotter BB component changes its luminosity by a factor 2.
This means that the BB radius of the hotter component is sensibly lower
during TOO2 BE.

The PE, with decreasing intensity, has been detected up to 10 keV in all but 
the last TOO observations (see Table~2). Its spectrum is well fit by a 2BB
model over the luminosity range over which the source could be studied
with the LECS and the MECS (from TOO1 to TOO3). The temperatures of the
two BB components do not change with time, except for a marginal evidence
of an increase in the hotter component during TOO3. The BB luminosities
instead do change. As a consequence, both BB radii (see Table~2) decrease 
with time: $R_1^{\rm BB}$ from 6.5 to 1.7~km, and $R_2^{\rm BB}$ from 0.8 to
0.2~km. 

Within the errors, PE and BE have hotter BB components with the same
temperatures, while the temperature of the cooler BB component is slightly
higher during PE than during BE (0.63$\pm$0.03~keV against 0.43$^{+0.04}_
{-0.03}$~keV).
During TOO1, the BB luminosities are however much different in the two
source states: the BE/PE luminosity ratio of the hotter BB is 64$\pm$5
during TOO1 and 32$\pm$2 in TOO2, while that of the colder BB is
18$\pm$7 in TOO1 and 44$\pm$12 during TOO2. Also, in TOO1, during the BE
the hotter BB component is brighter (by nearly a factor 3) than the cooler
BB, while during the PE they have similar luminosities. In TOO2, instead,
the latter ratio becomes $\sim$2 due to the lower PE luminosity of the
cooler BB.

On the basis of these features, a possible picture of the 2BB model
(already discussed by \cite{Guerriero99}) is that the colder component
originates from an accretion disk, while the hotter from the NS surface. 
A comptonizing plasma cloud is the likely origin of the BE high energy
component. The increase in luminosity of the colder BB component during
the BE can be interpreted as an increase of the disk emitting surface (by
a factor $\sim20$ in TOO1 and $\sim40$ in TOO2) as a consequence of the
reduction of the inner radius of the accretion disk, while that of the
hotter component could be due to an expansion in size (by a factor
$\sim60$ in TOO1 and $\sim30$ in TOO2) of the NS surface that emits
X--rays.

\begin{figure}
\epsfig{figure=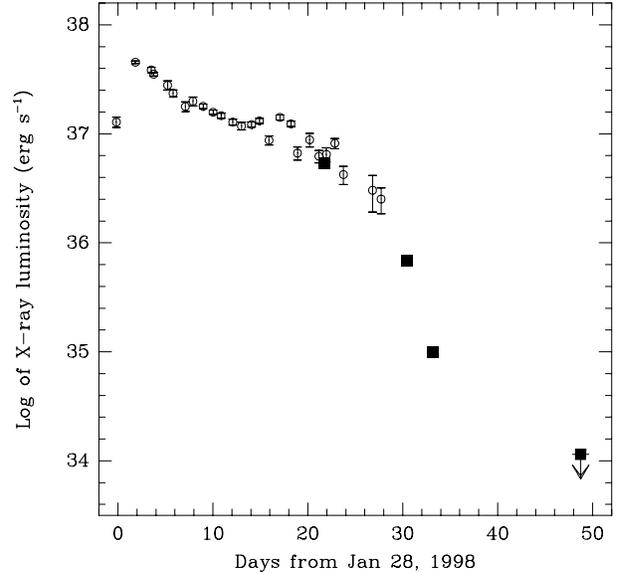,width=10cm}
\vspace{-1.5cm}
\caption[]{Light curve of the 1998 February-March outburst of the RB.
The filled squares are the {\it BeppoSAX} 2--10 keV fluxes derived using
the best-fit model given in Table~2, while the open circles represent
3-$\sigma$ detections of the RB taken from the {\it R-XTE}/ASM archive.
The errors of the {\it BeppoSAX} measurements were not plotted since they
are smaller than the corresponding symbols}
\end{figure}

\subsection{Outburst decay toward quiescence}
\label{lc}

Another relevant result of our observations is the determination, for the
first time, of the source flux decay toward quiescence. In Fig.~8 we show
the decay curve of the 1998 Feb-Mar outburst of the RB based on the 
{\it R-XTE}/ASM and \B\ data. As it can be seen, starting from day 28
after the
outburst onset, the flux decay becomes much faster than before: the
$e$-folding decay time $\tau$ changes from $\tau_1 \sim10$~days to
$\tau_2 \sim0.2$~days.
This behaviour is reminiscent of the final evolution of dwarf novae
outbursts (\cite{Osaki96}) and of the transition to quiescence of the SXTs 
that harbour a low magnetic field NS 
(e.g., Aql X-1, \cite{Campana98}; SAX J1808.4-3658, \cite{Gilfanov98}) 
or a black hole (see the
review by Tanaka \& Shibazaki 1996\nocite{Tanaka96}). In the case of
Aql X-1 (\cite{Campana98}) this time behaviour was interpreted as
a consequence of the propeller effect (\cite{Illarionov75}).
In the present case, the same effect can explain our observational
results also.
Indeed a magnetosphere may be present in the RB as consequence of
the relaxation oscillator character of the Type II bursts recurrence
behaviour (\cite{Lamb77}; Baan 1977, 1979) and low
frequency QPOs detected in the BE and PE (\cite{Stella88};
\cite{Lewin95}). In the model by Baan (1977, 1979), a reservoir of
matter floats on the top of the NS magnetosphere, partly supported
by centrifugal forces. In the intervals between bursts, the magnetosphere
becomes larger than the corotation radius. As matter accumulates on the
magnetospheric boundary, this shrinks. When the magnetospheric radius
becomes lower than a critical value at which gravity overcomes the
centrifugal and
magnetospheric forces, Type II bursts occur. This model is qualitatively 
in agreement with our observations. 
As long as we observed (in TOO1) Type II bursts, the flux decay was
on the extrapolation of the earlier light curve. During TOO2, when we
mainly observed PE (only 5 bursts at the beginning of the \B\ pointing
were detected), we found the first deviation. The sharper decrease of the
PE intensity continued down to our quiescent luminosity upper limit.

In this scenario, the absence of Type II bursts in this decay phase is due
to no accumulation of matter on the magnetospheric boundary, as
consequence of a low mass accretion  from the companion star.
This fact leads to an increased magnetospheric radius. When the
magnetospheric radius $R_m$ $=\, 9.8\times 10^{5} \mu_{26}^{4/7} M_{1.4}^
{1/7}(L_{38}R_6)^{-2/7}$~cm ($\mu_{26}$ is the magnetic dipole
moment in units of $10^{26}$~Gauss cm$^3$ and $M_{1.4}$ is the NS mass in
units of 1.4 $M_\odot$) becomes equal to the corotation radius $R_{cor}$
 $=\,1.67 \times 10^8 P^{2/3} M_{1.4}^{1/3}$~cm ($P$ is the spin
period), accretion is inhibited (\cite{Stella86}).
The corresponding minimum luminosity is given by
$L_{\rm min}\sim 4\times 10^{36}B_8^2 P_{-3}^{-7/3}\,$erg s$^{-1}$, where
$B\,=\,B_8 10^8$~Gauss is the surface magnetic field of the NS, and
$P\,=\,P_{-3}10^{-3}$~ms is the NS spin period (NS mass $M$ and radius
$R$ are assumed to be 1.4$M_\odot$ and 10$^6$~cm, respectively).
Once accretion onto the NS surface is inhibited, as also discussed by
Stella et al. (1994)\nocite{Stella94} and Campana et al. 
(1998)\nocite{Campana98}, a lower
accretion luminosity $L_{\rm cor}$, corresponding to a gravitational
potential energy of $GM\dot{M}_{\rm min}/R_{\rm cor}$, is released. It can
be easily shown that $L_{\rm cor}\sim 2\times
10^{36}B_8^2P_{-3}^{-3}\,$erg s$^{-1}$, with the same assumptions made for
$L_{\rm min}$ (\cite{Campana98}).
From Fig.~8, we can estimate $L_{\rm min} \sim 3\times 10^{36}$~erg s$^{-1}$.

On the other hand, Fox \& Lewin (1999)\nocite{Fox99}, from a timing
analysis of {\it R-XTE} data from the RB, found two peaks in the power
spectral density spectrum of the Type I BE, one at 154.9$\pm$0.1~Hz
and the other at 306.6$\pm$0.1~Hz.
Assuming the lower frequency as due to the spin period, from the estimated
value of $L_{\rm min}$ we would obtain $L_{\rm cor} \sim 6 \times 10
^{35}$~erg s$^{-1}$. This is significantly lower than the luminosity
measured during TOO2, which can be considered as a lower limit for
$L_{\rm cor}$.
A higher value of $L_{\rm cor}$ (9$\times 10^{35}$~erg s$^{-1}$)
is expected in the case we assume as spin period that corresponding to the
higher frequency reported by Fox \& Lewin (1999). This spin period (3.3
ms) is more consistent with our results and implies a magnetic field
$B \sim 4\times 10^8$ Gauss, which then appears, on the basis of our
luminosity data, the most likely value of $B$ to be associated to the NS
harboured in the system.

The contamination from 4U\thinspace1728--34 prevented us
to draw conclusions on the quiescent luminosity from the RB. Our 3$\sigma$
upper limit to this luminosity (1$\times$10$^{34}$ erg s$^{-1}$)
is consistent with the luminosity level (1.9$^{+1.2}_{-0.6} \times
10^{33}\,$ erg s$^{-1}$) reported by Asai et al.
(1996) during quiescence. The latter value is consistent with the minimum
accretion luminosity (1.6$\times$10$^{33}$ erg s$^{-1}$) that can be
emitted in the propeller regime ($\sim 2\times 10^{34}B_8^2P_{-3}^
{-9/2}\,$erg s$^{-1}$, \cite{Campana98}), even though the detected
luminosity, as discussed by Asai et al. (1996), could be partially due to
low-luminosity X--ray sources within the globular cluster Liller 1. An
unbiased estimate of the RB quiescent luminosity is highly desirable to
test the role of the propeller effect in quiescence.

\begin{acknowledgements}
We are grateful to P. Giommi for his help in the use of {\sc saxsim}
for the analysis of the TOO4 MECS data. {\it BeppoSAX} is a joint Italian
and Dutch programme. This research was partly supported by the Italian
Space Agency. 
\end{acknowledgements}

\end{document}